\documentclass[preprint]{aastex631}

\usepackage{natbib}
\usepackage{amsmath}
\usepackage{hyperref}
\usepackage{soul}

\renewcommand{\L}{$\hat{\text{\bf{L}}}$}
\newcommand{\M}{$\hat{\text{\bf{M}}}$}

\newcommand{\LM}{$\hat{\text{\bf{LM}}}$}

\begin{document}

\title{Current Sheets and Magnetic Reconnection statistics by Solar Orbiter at few to tens of Ion Inertial Lengths II: Conditions for Reconnection, and Current Sheet Planarity.}

\author{Inmaculada F. Albert}
\affiliation{Department of Electromagnetism and Electronics, University of Murcia, Murcia, Spain}

\author{Sergio Toledo-Redondo}
\affiliation{Department of Electromagnetism and Electronics, University of Murcia, Murcia, Spain}

\author{Víctor Montagud-Camps}
\affiliation{Departament de Física, Universitat de les Illes Balears, 07122 Palma de Mallorca, Spain}
\affiliation{Institute of Applied Computing \& Comunity Code (IAC3), UIB, Spain}

\author{Aida Castilla}
\affiliation{Department of Electromagnetism and Electronics, University of Murcia, Murcia, Spain}

\author{Naïs Fargette}
\affiliation{Department of Physics, Imperial College London, London, UK} 
\affiliation{Institut de Recherche en Astrophysique et Planétologie, CNES, Toulouse, France}

\author{Benoît Lavraud}
\affiliation{Laboratoire d’astrophysique de Bordeaux, Univ. Bordeaux, CNRS}
\affiliation{Institut de Recherche en Astrophysique et Planétologie, CNrS, Toulouse, France}

\author{Philippe Louarn}
\affiliation{Institut de Recherche en Astrophysique et Planétologie, CNRS, Toulouse, France}

\author{Christopher J. Owen}
\affiliation{Mullard Space Science Laboratory, University College London, Holmbury St. Mary, Dorking, Surrey, RH5 6NT, UK}

\author{Ioannis Zouganelis}
\affiliation{European Space Agency (ESA), European Space Astronomy Centre (ESAC), Villanueva de
la Cañada, Madrid, Spain}


%






\begin{abstract}
We present a statistical study of magnetic reconnection and current sheet (CS) properties at scales from the ion inertial length ($\delta_i$), up to 86.4 $\delta_i$. We analyze 103 hours of high-cadence (4 samples/s) in-situ Solar Orbiter ion data, collected from July 2020 to September 2022, which correspond mainly to slow and Alfv\'enic solar wind. We applied an automatic detection algorithm to find CSs of thicknesses near and above the ion inertial length ($\delta_i$), and classify them as Alfv\'enic or reconnecting. The orientation of the CSs is consistent with an isotropic distribution of the discontinuities. The median ratio of reconnection jet velocity to theoretical outflow velocity is 0.71. We find that the diamagnetic drift reconnection suppression mechanism has no impact on the cataloged CSs, due to the symmetry of their upstream conditions. We find that a flow shear in the reconnection direction suppresses magnetic reconnection, and it is consistently greater for Alfv\'enic than for reconnecting CSs.  Flow shear in the guide-field direction is too small to suppress magnetic reconnection in ion scale solar wind CSs.

  
\end{abstract}



\section{Introduction}\label{sec:intro}

Magnetic reconnection is a physical process by which the energy contained in the magnetic fields is released into the plasma, accelerating and heating it, while reconfiguring the topology of the magnetic field. Magnetic reconnection is closely related to turbulence: magnetic reconnection arises from turbulence and generates it. \citep[e.g][]{servidio2011magnetic, phan2018electron, matthaeus1986turbulent, ergun2018magnetic}. In turbulence, energy is transferred to increasingly small scales, eventually dissipating into particle acceleration and plasma heating. The rate at which turbulence dissipates energy is increased at scales of the order of one ion inertial length ($\delta_i$), where the spectral break is observed. Magnetic reconnection has been proposed as one of the mechanisms for this dissipation of energy \citep{cerri2017reconnection, rueda2021three, servidio2009magnetic, stawarz2024, franci2017magnetic, perrone2017coherent, papini2018}.In the companion to this paper \citep{Albert2026P1} we address the question of how much energy is dissipated via magnetic reconnection at scales of the ion inertial length.  
Further discussion of the interplay between magnetic reconnection and turbulence can be found in the introduction to that companion paper. In that work, we have shown that current sheets (CSs) of lengths between $\sim1$ and $\sim90$ $\delta_i$ in the solar wind are probably generated locally by turbulence. We also estimated that $\sim30$ J kg$^{-1}$ s$^{-1}$ are dissipated by magnetic reconnection at scales between $\sim 1 - 40$ $\delta_i$.

CSs have been observed in the solar wind since early space missions, including Explorer 43 \citep{burlaga1977interplanetary} and Pioneer \citep{tsurutani1979interplanetary}. Efforts to study CSs as directional discontinuities \citep{burlaga1969DD} and classify them as either tangential (TD) or rotational discontinuities (RD) have continued ever since \citep[e.g.][]{neugebauer1984RDTDearly, horbury2001RDTD, neugebauer2006RDTD, liu2023RDTD}. However, no clear proportion of RD to TD has yet been determined. 

Magnetic reconnection can take place within CSs, as identified for the first time in the solar wind by \citet{gosling2005}. Magnetic reconnection has been reported at a wide range of length-scales in the solar wind, including,  reconnection occurring at the heliospheric CS \citep[e.g.][]{gosling2006HCSrec, lavraud2009observation, phan2021HCSrec}, at coronal mass ejection boundaries \citep[e.g.][]{ruffenach2012multispacecraft, ruffenach2015statistical} and reconnection taking place in CSs of a few ion inertial lengths in thickness.
 
Several recent studies have focused on the search for small-scale magnetic reconnection in the solar wind, in the 1-10$^3$ $\delta_2$ scale-range. These works \citep{lavraud2021magnetic, fargette2023clustering, eriksson2022,eriksson2024parker,vasko2021Rec,Lotekar2022,vasko2024} are varied in methodology, sampled heliosphere region, and resolved scale range, but quantify the abundance of magnetic reconnection in the range of 0.1 - 10 exhausts per hour of observation. 
It has been observed that in the solar wind only a small fraction of CSs reconnect \citep{osman2014magnetic}. We aim to investigate the reason for this observation and why magnetic reconnection occurs in some CSs and not in others. To this end, we have developed a CS and magnetic reconnection detection method that is able to combine a small-scale resolution, exhaustiveness in the CS search, and the most robust method for magnetic reconnection determination, i.e. observation of reconnection jets in ion velocity. The method \citep{Albert2026P1} produces a catalog of reconnecting and non-reconnecting CSs in the thickness range from a few to $\sim 90$ $\delta_i$ that we can compare to better understand the necessary conditions for magnetic reconnection to take place. Several mechanisms that suppress the onset of magnetic reconnection have been proposed in the literature, in this article we asses the ability of each of theses mechanisms to impact the proportion of CS that reconnect in the solar wind.

\citet{swisdak2003diamagnetic} first proposed that the diamagnetic drift of the X-line may suppress magnetic reconnection. In asymmetric magnetic reconnection with non-zero guide field, the gradient in plasma pressure can cause a diamagnetic drift velocity of the X-line. When the drift velocity of the X-line approaches the Alfvén speed the reconnection outflow may be weakened or completely unable to develop. Therefore, according to \citet{swisdak2010vector}, magnetic reconnection can only proceed when the difference in plasma beta between the two inflow regions fulfills that:
\begin{equation}
    \Delta\beta<\frac{2L}{\delta_i}\tan{\left(\frac{\alpha_B}{2}\right)},
\end{equation}\label{eq:diamagnetic1}

where L is the characteristic thickness of the CS near the X-line, and $\alpha_B$ is the magnetic shear angle.
The ability of large diamagnetic drifts to suppress reconnection has been verified both using simulations \citep[e.g.][]{liu2016suppression} and observations \citep[e.g.][]{phan2010dependence,maheshwari2022investigation}. 

The presence of high velocity shears in the reconnection field (\L-shear) and guide field (\M-shear) directions has also been proposed as a factor that limits magnetic reconnection. Flow shear in the reconnection direction has been thoroughly studied since it was first proposed \citep{chen1990resistive} and has been shown to suppress magnetic reconnection \citep[e.g.][]{doss2015asymmetric}. \citet{cassakOtto2011} have  shown theoretically that the presence of a flow shear in the reconnection direction can modify the dimensional reconnection rate according to:
\begin{equation}
    E_{shear}=E_0\left(1-\left(\frac{\Delta v_L/2}{ v_{th}}\right)^2\right),
\end{equation}\label{eq:RecRate}

where $E_{shear}$ and $E_0$ stand for the dimensional reconnection rate in the presence and absence of a flow shear, respectively, $\Delta v_L$ is the \L-shear and $v_{th}$ is the theoretical outflow velocity of a hypothetical reconnection jet. Consequently, the reconnection rate becomes zero when the flow shear is equal to the Alfvén speed shear, which imposes a hard limit on the flow shears that can support magnetic reconnection, which must fulfill $\Delta v_L\leq2 v_{th}$. 
Recent observational studies \citep{phan2020parker, eriksson2022} rarely find reconnection events in which \L-shear is greater than half of that limit. 

In recent studies \citet{mallet2025suppression, mallet2025effect} delve into one of the processes through which flow shears suppress magnetic reconnection. The authors explore how the presence of flow shear affects the growth rate of the tearing mode instability in collisionless and resistive magnetic reconnection, respectively. The growth rate of the tearing mode is shown to decrease with increasing flow shear, with different functional dependencies for the two regimes. In particular, \citet{mallet2025suppression} shows that the growth rate of the tearing mode instability in collisionless plasmas 
is modified by the presence of a flow shear as:
\begin{equation}
    \frac{\gamma_{tr}}{\gamma_{0,tr}}=\frac{1-\Sigma^2}{1+\Sigma^2\tau Z},
\end{equation} \label{eq:TearingGrowth}

where $\tau=T_i/T_e$, $Z=q_i/e$, and $\Sigma=\delta v/\delta b$, is the ratio of ion velocity fluctuations ($\delta v)$ to Alfvén speed fluctuations ($\delta b$). The decrease in the growth rate of this instability precludes the CS from initiating magnetic reconnection.

 \citet{ervin2025impact} has been able to link the significant decrease in the growth rate of the collisionless tearing instability due to large flow shears of \citet{mallet2025suppression} to a solar wind regime of high alfvénicity and slow bulk flow. They show how observations of the near-Sun solar wind periods containing reconnecting CSs systematically present smaller flow shears and larger collisionless tearing instability growth rate than periods with no reconnection. Additionally, they show that 85\% of reconnection periods occur in non Alfvénic slow solar wind environments.

The effects of guide field-directed flow shear have received less attention. \citet{wang2008out} showed in simulations how the presence of a flow shear in the guide field direction generates a quadrupolar out-of-plane magnetic field signature, similar to the quadrupolar Hall field signature of magnetic reconnection. Simulations by \citet{chen2013influence} demonstrated that flow shears in the direction of the guide field can modify magnetic reconnection,  generating a secondary tearing instability that leads to the formation of secondary X lines and magnetic islands. The effects of flow-shear on the ability of magnetic reconnection have been more recently studied in simulations.  \citet{liang2024impact} find that the presence of a guide field-directed flow shear can enhance the reconnection rate and energy conversion in magnetopause reconnection. \citet{swisdak2026numerical} show in simulations how strong flow shears in the guide field direction can decrease the reconnection rate and increase the onset time of the tearing mode instability, thus leading to the suppression of magnetic reconnection. In this study, we will assess the importance of all these mechanisms in the suppression or enablement of magnetic reconnection at scales in the range of between 1 $\delta_i$ and $\sim 90$ $ \delta_i$ in the solar wind.

Regardless of whether magnetic reconnection takes place in them, the structure of CSs is usually assumed to be bi-dimensional, following the paradigm of the Sweet-Parker and Petschek models \citep{sweet1958electromagnetic, parker1957acceleration, petschek1964magnetic, kulsrud2001magnetic}, in which CSs are assumed to occur in a flat, infinite 1D plane. However, magnetic reconnection rarely occurs in such a perfect configuration, and is usually three dimensional \citep[e.g.][]{ji2022magnetic,daughton2014computing}. In this work, we investigate the  planarity of the small-scale CSs that arise in the solar wind, and how much they deviate from the flat planar picture.

The present paper is structured as follows. In section \ref{sec:data} we introduce the dataset in which the study is conducted and provide an overview of the algorithm used for the automatic detection of CS and the assessment of reconnection there. In section \ref{sec:orientation} we study the orientation of the CSs in the catalog. In section \ref{sec:RDTD} we discuss the nature of CSs in terms of types of discontinuities. In section  \ref{sec:Suppression} we assess whether magnetic reconnection is suppressed in the sampled solar wind environment, and present results on the Alfvénicity degree of CSs, contrasting reconnecting and non-reconnecting CSs. In section \ref{sec:flatness} we discuss the planarity of the presented CSs.

\color{black}
\section{Methodology}\label{sec:data}


We use Solar Orbiter (SolO, \citet{Muller2020SolO})
data from the Magnetometer (MAG, \citet{Horbury2020MAG}) and the Proton Alpha Sensor (PAS, \citet{Louarn2021PAS}) of the solar wind Analyser (SWA, \citet{Owen2020SWA}) instruments, collected between July 2020 and September 2022. We select data intervals with high resolution, i.e., 4 vectors per second, of the ion Velocity Distribution Function (VDF), as we aim to find ion-scale structures. Our dataset comprises 103 hours of data, distributed in 25,226 separate intervals, which constitute our Dwell Time (DT). 24,615 of these intervals are 8 s long and amount to 54.7 hours. The rest of the DT intervals are 5 minutes long.  Typical parameters of slow solar wind near 1 AU are an ion velocity of $v\sim400$ km/s, a magnetic field magnitude of $B_0\sim 5$ nT, and a plasma density of $n\sim$ 5 cm$^{-3}$, resulting in an ion inertial length of $\delta_i\sim$ 100 km. 

The Solar Orbiter heliocentric distance during the DT ranges from 0.4 AU to 1 AU. The measured solar wind velocity ranges from 280 km/s to 770 km/s, with 94\% of the DT being slow solar wind, i.e., below 500 km/s. The solar wind during the DT presents a median Mach number 10.4, with the tenth and ninetieth percentiles being 5.7 and 16.7 respectively. The solar wind magnetic field and ion number density first and third quartiles are 3.8 nT and 6.5 nT and 6.2 cm$^{-3}$ and 12.7 cm$^{-3}$, respectively. For more details on the dataset, please see the companion paper, \citet{Albert2026P1}. 

We used the collected dataset described above as a high resolution sampling of the solar wind and developed an algorithm to search for CSs of characteristic width close to ion length-scales. 
We present a catalog of CSs of crossing lengths in the range $\Lambda\in[v_{SW}/(1/\Delta t_{PAS}+1/\Delta t_{MAG}), 100\delta_i] $ where the magnetic field rotates monotonically and the rotation exceeds a minimum threshold. This threshold is defined for each data segment as $\alpha_B\ge 2\arcsin{\left(\frac{\sqrt{\mu_0n_im_i}}{|\mathbf{B}|}\Delta v_{PAS}\right)}$. This ensures that if a velocity jet were present in the CS, it would be detectable above the velocity uncertainty level. We take the value $\Delta v_{PAS}=10$ km/s, three times larger than the uncertainty of PAS as provided by \citet{Louarn2021PAS}. We require a reversal in the $\hat{\bf{L}}$ component of the magnetic field in the reference system built by hybrid Minimum Variance Analysis \citep{gosling2013magnetic}.
When several CS candidates overlap, the candidate that presents the maximum magnetic field rotation is selected.

\begin{figure*}[hb]
\includegraphics[width=0.95\textwidth]{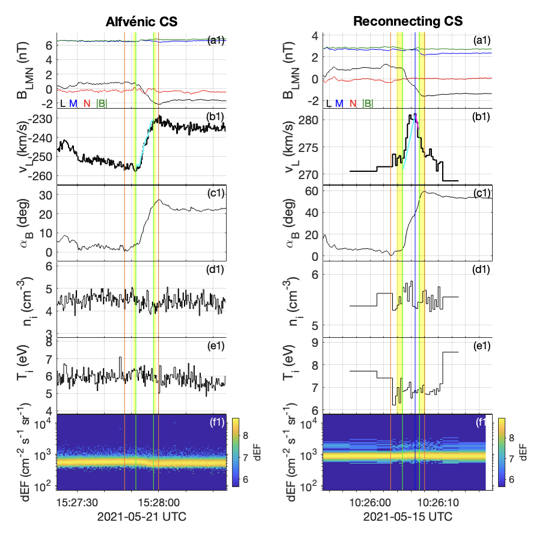}
\caption{Details of two CSs found by the algorithm. The left panels (labeled 1) show an example of an Alfvénic CS. The right panels (labeled 2) show an example of a Reconnecting CS. (a) Magnetic field components and magnitude in LMN coordinates using Hybrid MVA. (b) Ion velocity in the $L$ direction (black), Alfvén velocity increments in the $L$ direction correlated and anti-correlated to the magnetic field (magenta and cyan, respectively). (c) Angle rotated by the magnetic field. (d) SWA-PAS ion number density. (e) SWA-PAS ion scalar temperature. (f) Ion differential energy flux (in keV/(cm$^2$ s sr keV)). The vertical orange lines show the algorithmically selected CS, encompassing 100\% of the rotation, while the green lines show the central 76\% rotation of the CSs. The blue vertical line marks the center of the jet, defined as the peak in the velocity beam. The yellow shaded areas encompass the three datapoints immediately outside each side of the central 76\% rotation region. The Alfvénic CS shown in the left panels is captured during a 5 min data segment. The Reconnecting CS pictured in the right panels is detected during a short 8 s data segment. We have only shown the ion data captured in snapshot mode of SWA-PAS. At the two sides of the Reconnecting CS detection, plasma data are collected in the normal mode of SWA-PAS, however aberrations in the data appear when the cadence of data collection is changed.
\label{fig:CSplot}}
\end{figure*}

The found CSs are classified as "Alfvénic", "Reconnecting", or "Undetermined", depending on the correlation coefficient between the magnetic field and ion velocity in the leading and trailing halves of the CS. If the correlation coefficient is larger than 0.5 and has the same sign in the two halves of the CS, it is labeled as "Alfvénic". If the correlation coefficient is larger than 0.5 in the two halves of the CS but has opposite signs, it is labeled as "Reconnecting". If the correlation coefficient is smaller than 0.5 in either half, the CS is labeled as "Undetermined".

Figure \ref{fig:CSplot} shows examples of two CSs, one Alfvénic (left, 1) and one reconnecting (right, 2), found by the algorithm described above. The figure shows the time evolution of the magnetic field (a), ion velocity (b), magnetic shear angle (c), ion density (d), ion temperature (e) and the ion spectrogram (f) during the two events.

\section{Results and Discussion}\label{sec:results}

Applying the algorithm explained above to the 103 hours of dwell time, provides a catalog of 974 CSs of normal width $0.05$ $\delta_i$ $\le\lambda\le 86.4$ $\delta_i$. Of them,  838 are Alfvénic, 83 are undetermined and 53 are reconnecting, which amounts to 5\% of CSs undergoing magnetic reconnection. We find that, in total, Solar Orbiter spends  4,390 s 
 inside a CS and 290 s 
 inside a reconnecting CS, that is, 1.2 
 and 0.08 
percent of the dwell time, respectively. 
 A more in-depth discussion of the CS size can be found in the companion article \citet{Albert2026P1}, specifically in Figure 3 and Table 1.
The CS in this catalog are classified as "Alfvénic" if they present a high correlation between the magnetic field and velocity in the  $\hat{\text{\bf{L}}}$ direction, as "Reconnecting" if the correlation between magnetic field and velocity in the  $\hat{\text{\bf{L}}}$ direction is high throughout the CS but changes sign at some point, and as "Undetermined" if the correlation between the magnitudes is low.

\subsection{Current sheet orientation}\label{sec:orientation}

\begin{figure}[ht]
\epsscale{1.15}
\plotone{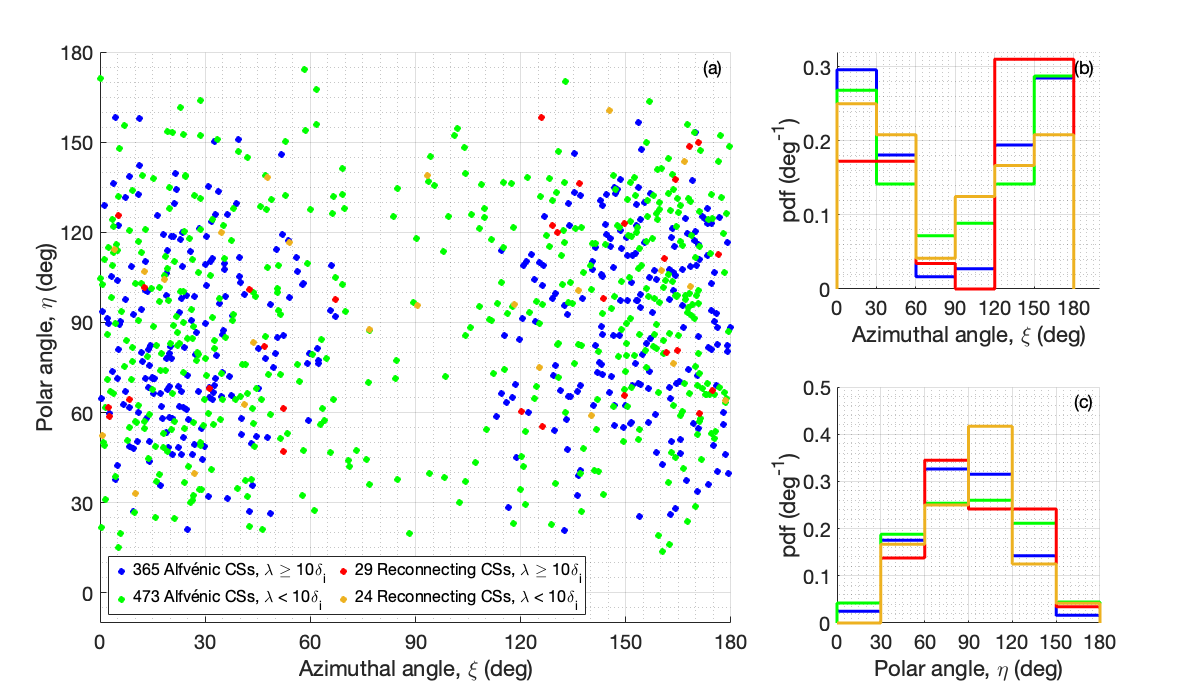}
\caption{Each dot in panel (a) represents the azimuthal and polar angles of the  $\hat{\textbf{N}}$ of a CS in the catalog. Panels (b) and (c) present the distribution of the same azimuthal and polar angles, respectively. The four colors blue, green, red and yellow represent Alfvénic CSs of thickness $\lambda\geq10\delta_i$ (365), Alfvénic CSs of thickness $\lambda<10\delta_i$ (473), reconnecting CSs of thickness $\lambda\geq10\delta_i$ (29), and reconnecting CSs of thickness $\lambda<10\delta_i$ (24), respectively.
\label{fig:Orientation}}
\end{figure}

The orientation of the normal vector of each CS in the RTN coordinate system is shown in Figure \ref{fig:Orientation}, where the polar and azimuthal angle are defined as $\eta=\arccos{(N_z/|\hat{\text{\bf{N}}}|)}$ and $\xi=\text{sign}({N_y})\arccos{\left(N_x/\sqrt{N_x^2+N_y^2}\right)}$. In this coordinate system $\eta=0^{\circ}$ corresponds to a $\hat{\text{\bf{N}}}$ vector that is directed northward, out of the ecliptic plane; $\eta=90^{\circ}$, $\xi=0^{\circ}$ corresponds to a normal vector that is in the radial direction; and   $\eta=90^{\circ}$, $\xi=90^{\circ}$ corresponds to a  $\hat{\text{\bf{N}}}$ that is along the tangential direction. The distribution of polar and azimuthal angles of the normal vectors of the CSs in the catalog presents two gaps around $\xi=90^{\circ}$ and $\eta=0^{\circ}$. This indicates a lack of CSs with  normal vectors that are approximately perpendicular to the radial direction, i. e., CSs that would be traversed by Solar Orbiter primarily along their $\hat{\text{\bf{M}}}$ or $\hat{\text{\bf{L}}}$. This behavior is consistent with a selection effect introduced by the observation method: if the trajectory of Solar Orbiter across a CS does not have a significant component along its $\hat{\text{\bf{N}}}$ direction, the  associated rotation of the magnetic field  cannot be observed, and the event may therefore not be identifiable as a CS crossing by our algorithm. Conversely, CSs sampled along directions that are nearly perpendicular to $\hat{\text{\bf{N}}}$ can still be detected and may allow for the detection of thinner CSs. In such cases,  the distance between consecutive data points corresponds to a smaller thickness, increasing the number of data points available to resolve the structure. 

Apart from the two gaps discussed above, no other features are apparent in the angle distributions. Thus, we conclude that the distribution is consistent with an isotropic distribution of the orientation of solar wind CSs in the 1 $\delta_i$ to 90 $\delta_i$ thickness range, the main deviations being attributable to geometric effects. All CSs follow the same distribution regardless of their thickness and reconnection status.

\subsection{Current sheets as discontinuities}\label{sec:RDTD}

CSs can be thought of as plasma discontinuities, separating two regions of plasma that may present changes in properties such as density, velocity, or magnetic field. Discontinuities can be classified as either rotational discontinuities (RDs) or tangential discontinuities (TDs) depending on the changes in plasma properties that take place across them. The properties of both types of discontinuity are described in more detail in \citet{BaumjohanTreumann}. According to their description, the magnetic field and velocity in RDs are non-zero and constant in the direction normal to the discontinuity, and the magnetic field magnitude is also constant. In contrast, TD are characterized by $B_N = 0$ and $v_N = 0$, with changes in the magnitudes of B and v from one side of the discontinuity to the other. RDs present a high correlation between the change in  magnetic field and velocity in the directions tangential to the discontinuity, so that the change in ion velocity from one side of the discontinuity to the other should should be equal to the change in Alfvén velocity in these directions, fulfilling $\Delta v_L =\pm\Delta v_{A,L}$. TDs can exhibit the same property, as non-zero changes in magnetic field and velocity across the discontinuity are expected, but are not required to.

 In this section, we assess whether each kind of CS (Reconnecting, Alfvénic, and Undetermined) corresponds to a particular kind of discontinuity. To that end, we will examine the jumps in the $\hat{\text{\bf{L}}}$-directed ion velocity and Alfvén velocity, and the jump in the magnetic field magnitude and the value of $B_N$ for each CS. In particular, Reconnecting CSs may present a bifurcated magnetic field, in which the magnetic field rotates in two steps, each corresponding to an individual discontinuity. Each Alfvénic and Undetermined CS can therefore be regarded as a discontinuity, while we analyze each reconnecting CS as a succession of two discontinuities occurring one immediately after the other. We delimit the two discontinuities associated with the two step rotation of a reconnecting CS at the center of the ion velocity jet (the point where the velocity data reaches an absolute maximum or minimum).
Thus, the change ($\Delta$) and average ($\langle\cdot\rangle$) of a physical quantity ($x$)  across a discontinuity are computed between the two sides of the discontinuity. For Alfvénic and Undetermined CSs, those two sides are the two sides of the CS, corresponding to the average value of the given quantity computed in the three points immediately exterior to the leading and trailing edges of the CS, marked by the yellow areas of Figure \ref{fig:CSplot}: $\Delta(x)=|x_l-x_t|$ and $\langle x\rangle$ is the average value of al the measurements of $x$ comprised between $x_l$ and $x_t$. In Reconnecting CSs two discontinuities are described, and the two are separated at the peak of the ion jet (marked by the blue vertical line in the right panels of Figure \ref{fig:CSplot}). The changes in a physical quantity across the two discontinuities associated with a Reconnecting CS are computed between one side of the CS and the point where the ion jet reaches its peak: $\Delta(x)=|x_l-x_{peak}|$ for the leading half and $\Delta(x)=|x_t-x_{peak}|$ for the trailing half. The averages of the quantities are computed as the average value of all the measurements comprised between $x_l$ and $x_{peak}$ and between $x_t$ and $x_{peak}$, respectively.

\begin{figure}[hb]
\epsscale{1.15}
\plotone{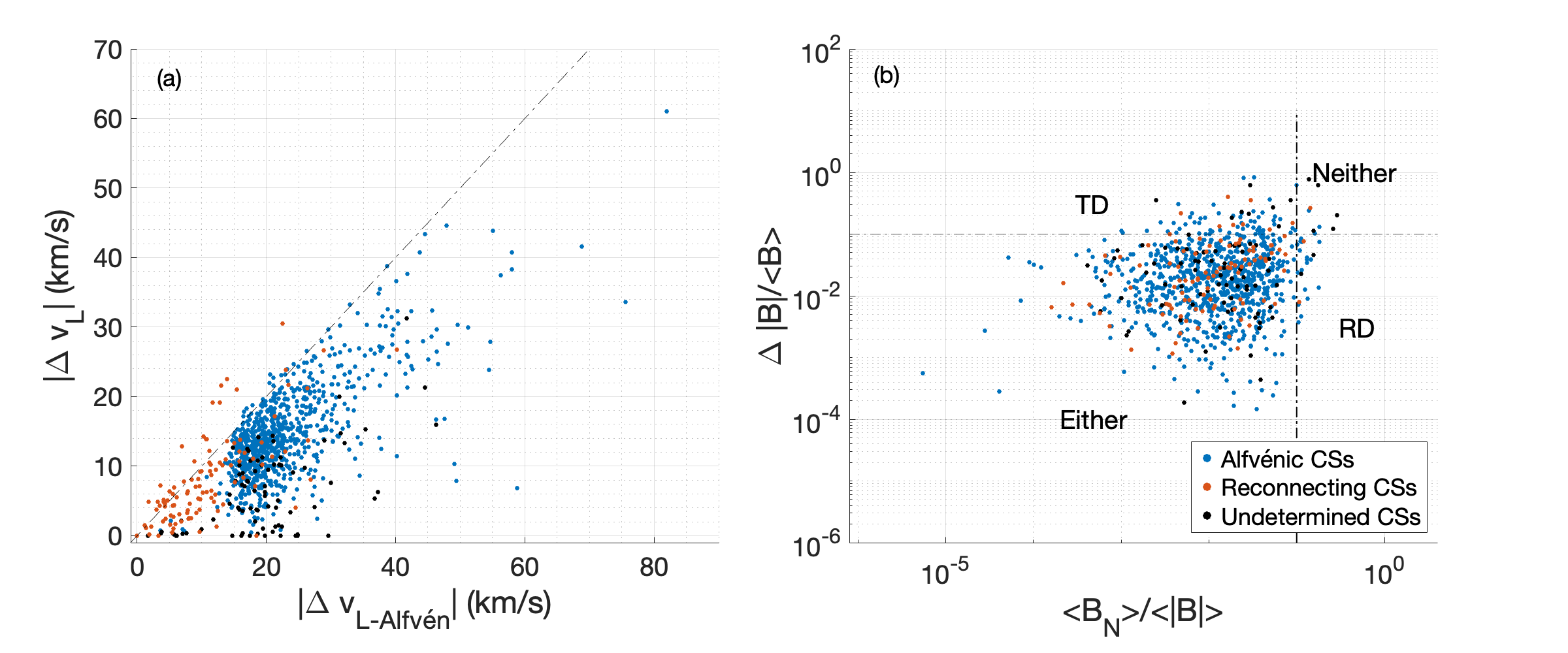}
\caption{Each red/blue/black dot in this figure represents a discontinuity associated with a Reconnecting/Alfvénic/Undetermined CS. Panel (a) shows the jump across the discontinuity in the $\hat{\text{\bf{L}}}$ component of the velocity versus the jump in the same component of the Alfvén velocity. Panel (b) shows the magnetic field magnitude jump across the discontinuity versus the normalized $\hat{\text{\bf{N}}}$ component of the magnetic field. The dash-dotted line in panel (a) is a visual aid representing where the absolute value of the jump in velocity equals the absolute value of the jump in Alfvén velocity. The dash-dotted lines in panel (b) represent $<B_N>/<|B|>=0.1$ (vertical) and $\Delta |B|/<|B|>=0.1$ (horizontal). These three lines represent the thresholds used to assess the type of discontinuity.  
\label{fig:RD_TD}}
\end{figure}

 In Figure \ref{fig:RD_TD} we present several properties of all the discontinuities associated with the CSs in our catalog. Figure \ref{fig:RD_TD}(a) shows the jump in the $\hat{\text{\bf{L}}}$ component of the ion and Alfvén velocities across each discontinuity. Figure \ref{fig:RD_TD}(b) shows the jump in magnetic field magnitude across the discontinuity versus the mean value of $B_N$ throughout the discontinuity. The two magnitudes are normalized to the mean magnetic field strength.  As a result, the CSs close to the diagonal dashed line in panel (a) of Figure \ref{fig:RD_TD} and in the bottom right quadrant of panel (b) are consistent with RDs, while the CSs in the top left quadrant of panel (b) are consistent with TDs, irrespective of their position in panel (a).  A large majority of CSs occupy the bottom left quadrant of panel (b), meaning that they may be either RDs or TDs, whereas a point in the top right indicates a structure that can be neither a RD nor a TD. We chose the thresholds delimiting the four quadrants to be 0.1, a value in the range used by previous authors \citep[e.g.][]{neugebauer2006RDTD, horbury2001RDTD, knetter2004RDTD, liu2023RDTD}. For a discontinuity to be classified as a RD, it must be in the bottom right quadrant of Figure \ref{fig:RD_TD}(b) and close to the $\Delta v_L = \pm \Delta v_{A,L}$ dashed line of Figure \ref{fig:RD_TD}(a). To be considered a TD, the discontinuity must be in the top left quadrant of Figure \ref{fig:RD_TD}(b).

After our analysis, 75 out of 1027 discontinuities (including two discontinuities per reconnecting CS) are classified as TDs (7.3\%), see Figure \ref{fig:RD_TD}(b). Nine of the 83 undetermined CSs are classified as TDs (11\%). Nine of the 106 discontinuities associated with reconnecting CSs are classified as TD (8.5\%). This is an unanticipated result,  reconnecting CS would not be expected to be associated with TDs, as this kind of discontinuity does not allow for the flow of plasma in the normal direction to the discontinuity, which would be necessary for magnetic reconnection to take place. However, the result is consistent with the study by \citet{servidio2011}, who observe that a fraction of the TDs identified in their simulation correspond to reconnection sites. 

Notably, we find even fewer confirmed RDs than TDs in our dataset. Twelve discontinuities are classified as RDs, while a vast majority of the discontinuities (91\%) are in the "either" category. These results are in stark contrast with those of \citet{liu2023RDTD} who finds a similar proportion of TDs, but RDs account for up to 75\% of their dataset. The comparative lack of RDs in our data set can be explained by two concurring reasons. First, the fact that RDs require $\Delta v_L=\pm\Delta v_{A,L}$ means that all Alfvénic CS should be classified as RDs, because of the strong correlation between magnetic field and velocity in the $\hat{\text{\bf{L}}}$ direction required to classify a CS as Alfvénic. This is not the case because although the condition is necessary for a discontinuity to be a RD, other kinds of discontinuity can also fulfill it, meaning that this property is insufficient to classify Alfvénic CSs as RDs on its own. Second, the use of hybrid-MVA to compute the LMN coordinate system of the CSs and the discontinuities minimizes the value of $B_N$ while maximizing $B_M$. This precludes Alfvénic CS classification as RDs, putting them in the "either" or TD categories. The same difficulty affects the classification of the discontinuities associated with reconnecting CSs. Moreover, the two discontinuities associated with reconnection exhausts are unlikely to fit the model of either a TD or n RD. It has been proposed that discontinuities at the edges of reconnection exhausts are a combination of a RD and a slow shock \citep{liu2012structure}.
Classification of solar wind CSs according to the type of discontinuity therefore yields inconclusive results when the hybrid-MVA method is applied. Nevertheless, hybrid MVA remains the preferred method for LMN frame computation in the solar wind, as it is the single spacecraft method that has best reproduced the results of spacecraft methods in the solar wind \citep{wang2024}. This result calls for an updating of the criteria that define TDs and RDs in the solar wind, in which the normal magnetic field may be underestimated.

  \color{black}

\subsection{Current sheet properties and magnetic reconnection suppression mechanisms.} \label{sec:Suppression}
We turn now to investigating some properties of these CSs.  We look at two conditions for magnetic reconnection suppression to determine which factors influence the proportion of reconnecting versus Alfvénic CSs: suppression by the diamagnetic drift of the X-line and by shear in ion velocity flows.

\begin{figure}[b]
\epsscale{1.15}
\plotone{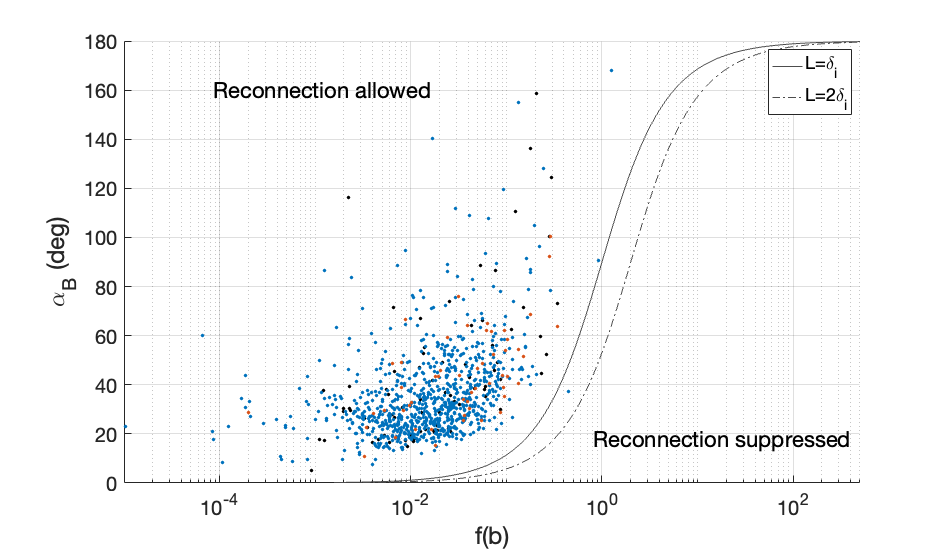}
\caption{Swisdak test. The magnetic field shear angle ($\alpha_B$) is shown against $f(b)=\ln(b^2) \sqrt{b}/(1+b)$, where $b$ is the ratio between the magnetic field magnitude on the two sides of the CS such that $b>1$. The blue, red and gray4 dots represent Alfvénic, reconnecting and undetermined CSs respectively. The solid and dot-dashed lines represent the limit where $f(b)=(L/\delta_i)\tan(\alpha_B/2)$ for $L=\delta_i$ and $L=2\delta_i$ respectively. L represents the characteristic thickness of the CS at the reconnection site. 
\label{fig:CSSwisdak}}
\end{figure}

We perform the Swisdak test for diamagnetic suppression of  magnetic reconnection \citep{swisdak2003diamagnetic,swisdak2010vector}. The asymmetry in plasma beta at the two inflow sides in a reconnection event causes a diamagnetic drift of the X-line in the outflow direction. If this drift is faster than the theoretical speed of the jet, reconnection cannot proceed. For any given event, for magnetic reconnection to be possible, the jump in plasma beta from one side to the other of the CS must satisfy Equation \ref{eq:diamagnetic1}. Considering that pressure balance must be maintained in the CS, an asymmetry in plasma beta implies an asymmetry in the magnetic field. Therefore, the previous condition can be rewritten in terms of the magnetic field, as in the study by \citet{eriksson2024parker}. For a CS to be able to reconnect, it must therefore satisfy the following:

\begin{equation}
    f(b) = \ln(b^2)\sqrt{b}/(1+b) > (L/\delta_i)\tan(\alpha_B/2),
\end{equation}
where $b=B_0/B_{100}$ is the ratio of the magnetic field strengths in either of the inflow regions of the CS. $B_0$ and $B_{100}$ represent the points where 0\% and 100\% of magnetic field rotation respectively have been completed.

In Figure \ref{fig:CSSwisdak}, all Reconnecting CSs from our catalog satisfy the condition, as they are located to the top left of the line representing the limit of the diamagnetic reconnection suppression condition. All the non-reconnecting CSs also meet the condition. This means that magnetic reconnection is not typically suppressed by the diamagnetic drift of the X-line under the plasma conditions sampled in this study. We attribute this to small-scale solar wind CSs being quite symmetric in magnetic field, meaning that they do not experience large diamagnetic drifts at the X-line that may hinder reconnection onset.
\color{black}

\begin{figure}[b]

\epsscale{1.15}
\plotone{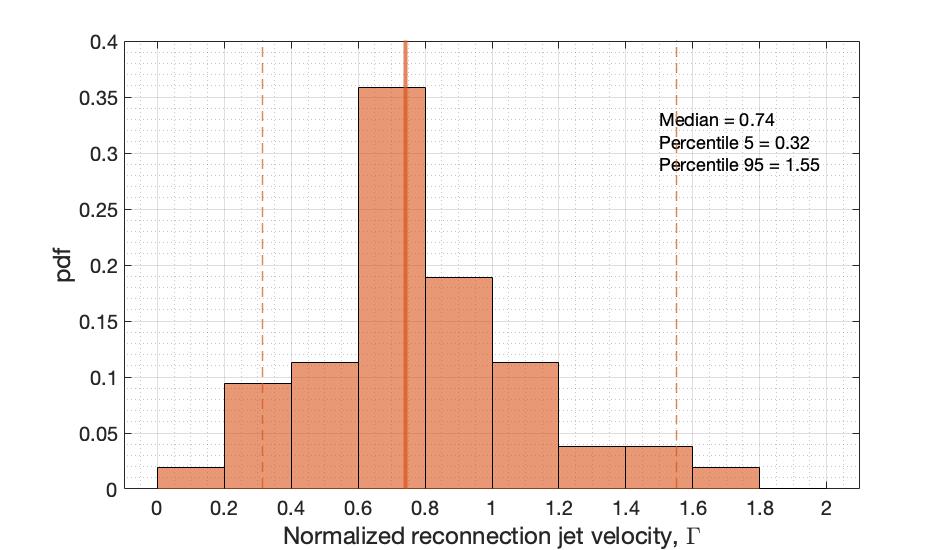}
\caption{Normalized jet velocity, $\Gamma$, of reconnecting CSs. The solid and dashed lines indicate the median and fifth and ninety-fifth percentiles of the distribution
\label{fig:CSalfvenicity}}
\end{figure}

Figure \ref{fig:CSalfvenicity} shows the normalized jet velocity, $\Gamma$, of the reconnecting CSs. The normalized jet velocity is the the velocity of each measured jet compared to the theoretically expected jet velocity ($v_L^{th}$). The jet velocity is computed as the difference between the velocity at the peak of the jet and the "baseline" velocity throughout the CS in the $\hat{\text{\bf{L}}}$ direction. For each edge of the CS, the environment velocity is computed as the mean velocity of the three datapoints immediately outside the CS (shown as the yellow shaded regions in Figure \ref{fig:CSplot}). We then define the "baseline" velocity as the average of the two environment velocities at each CS border. The measured velocity of the jet is then normalized to the theoretical jet velocity, as computed by \citet{CassakShay2007}, i.e., the Alfvén velocity in the $\hat{\text{\bf{L}}}$ direction for asymmetric reconnection

 \begin{equation} 
        \Gamma   = \frac{v_{L}}{v_{L}^{th}} =\frac{|v_{L}^{jet}-v_{L}^{baseline}|}{|v_{L}^{th}|},     
 \end{equation}\label{eq:JetAlfvenicity}
 
where $v^{th}_L$ is
\begin{equation}
        (v_L^{th})^2 =\frac{B_{l_L}B_{t_L}}{\mu_0}\frac{B_{l_L}+B_{t_L}}{\rho_l B_{t_L}+\rho_tB_{l_L}},
 \end{equation}

where the subscripts "$_l$" and "$_t$" stand for the leading and trailing edges of each CS.

 Reconnecting CSs usually have slower jets than the theoretical prediction, with a median normalized jet velocity of 0.74, and a distribution ranging from 0.06 to 1.8. These results show that the jets are faster than, but comparable to, those reported by \citet{eriksson2022}, who find that the median size of the reconnection jets is 0.6. However, their study comprises CSs up to one order of magnitude wider. 22\% of the jets are super-Alfvénic ($\Gamma>1$), which exceeds the findings of \citet{eriksson2022}. The large amount of Super-Alfvénic reconnection jets may be due to an inaccuracy in the computation of the normalized jet velocity coming from the uncertainty in the determination of the LMN coordinate system, or from some uncertainty in the computation of $v_L^{baseline}$ related to imprecision in CS border selection.

\begin{figure}[b]
\epsscale{1.15}
\plotone{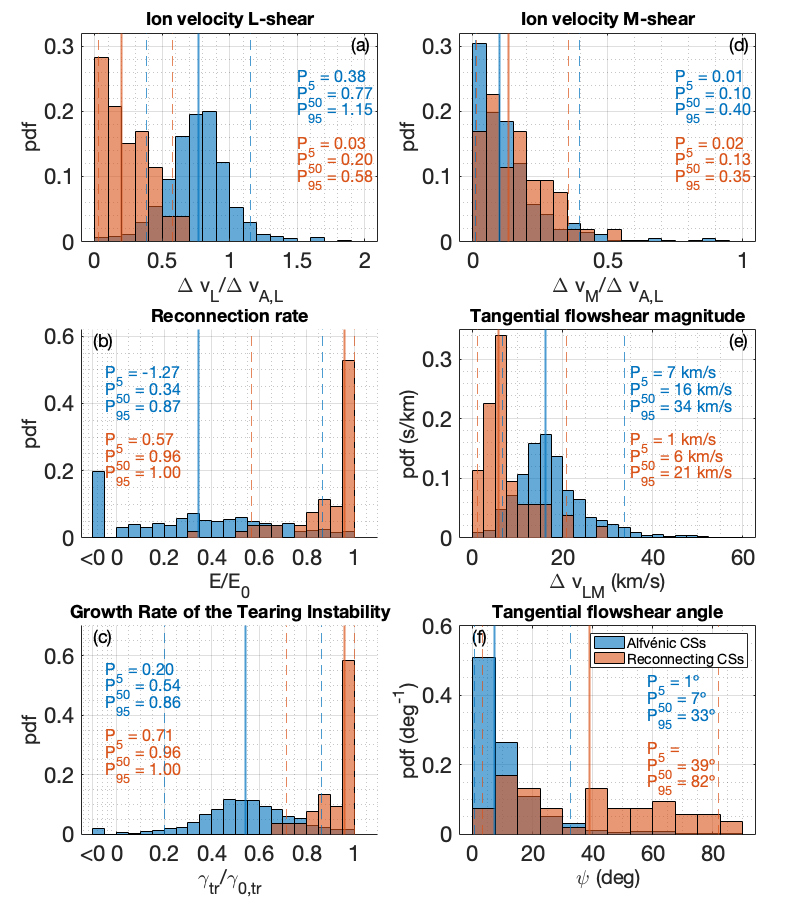}
\caption{(a) Normalized flow shear in the $\hat{\text{\bf{L}}}$ direction, defined as $|v_{L_l}-v_{L_t}|/|v_{A,L_l}-v_{A,L_t}| $. (b) Normalized reconnection rate as a function of \L-shear. (c) Normalized growth rate of the tearing instability as a function of \L-shear. (d) Normalized flow shear in the $\hat{\text{\bf{M}}}$ direction, defined as $|v_{M_l}-v_{M_t}|/|v_{A,M_l}-v_{A,M_t}| $. (e) Tangential flow shear magnitude. (f) Tangential flow shear $\psi$-angle. The vertical solid lines represent the medians of the distributions and the dashed vertical lines represent the fifth and ninety-fifth percentiles, P$_5$ and P$_{95}$, of the distributions. Blue and red distributions, lines and text correspond to Alfvénic and reconnecting CSs, respectively.
\label{fig:CSflowshear}}
\end{figure}

It has been proposed that shear ion flows in the $\hat{\text{\bf{L}}}$  and $\hat{\text{\bf{M}}}$ directions can affect the ability of magnetic reconnection to proceed. In particular, large \L-shears have been shown to suppress magnetic reconnection \citep[e.g.][]{cassakOtto2011,phan2020parker}, while some authors have suggested that a large \M-shears  may affect the growth-rate of the instability of the tearing mode, therefore affecting the onset of magnetic reconnection. In this regard, simulations by \citet{liang2024impact}, show the reconnection rate  increases in the presence of strong \M-shears in asymmetric CSs. Conversely, simulations by \citet{swisdak2026numerical} show the magnetic reconnection rate and the growth-rate of the tearing-mode instability decreasing with increasing \M-shear.  Figure \ref{fig:CSflowshear} shows the normalized flow shear of Alfvénic and reconnecting CSs, in the $\hat{\text{\bf{L}}}$  and $\hat{\text{\bf{M}}}$ directions, where the two flow shears are defined as :

\begin{equation}
\begin{split}   
    \Delta v_{L}= & |v_{L_l}-v_{L_t}|,\\
    \Delta v_{M}= & |v_{M_l}-v_{M_t}|,
\end{split}
\end{equation}

respectively, and are normalized to the shear in the $\hat{\text{\bf{L}}}$ component of the Alfvén ion velocity, $ \Delta v_{A,L}=|v_{A,L_l}-v_{A,L_l}|$, in accordance with \citet{phan2020parker, eriksson2022}, where the subscripts "$_{l}$" and "$_{t}$" denote the leading and trailing edges of the CSs, respectively. Please note that $\Delta v_L$, $\Delta v_M$ and $\Delta v_{A,L}$ are not frame dependent  under an inertial coordinate system transformation. The velocities at the edges are computed as the average of the velocity in the three data points immediately outside of the CS from each edge. Figure \ref{fig:CSflowshear} (e,f) also shows the magnitude,  $\Delta v_{LM}$, and orientation angle, $\psi$, of a tangential flow shear (\LM-shear) vector defined in the $\hat{\text{\bf{LM}}}$-plane respectively, i. e., the modulus of the vector formed by the velocity shears in the $\hat{\text{\bf{L}}}$  and $\hat{\text{\bf{M}}}$ directions, and the angle formed by that vector with the $\hat{\text{\bf{L}}}$ direction:

\begin{equation}
\begin{split}   
    \Delta v_{LM} = & \sqrt{\Delta v_{L}^2 + \Delta v_{M}^2},\\
    \psi = & \arctan{\left(\frac{\Delta v_{M}}{\Delta v_{L}}\right)},
\end{split}
\end{equation}

In Figure \ref{fig:CSflowshear}(a),  the normalized flow shear in the $\hat{\text{\bf{L}}}$ direction is significantly higher for Alfvénic than for reconnecting CSs: the median value of the normalized \L-shear is 0.71 for Alfvénic versus 0.20 for reconnecting CSs. Alfvénic CSs, which constitute a majority of the CSs we fin in the solar wind, can be regarded as sites where a magnetic field rotation coincides with a change of the $\hat{\text{\bf{L}}}$ component plasma velocity, and thus a non zero \L-shear is required by definition.

 The presence of a flow shear in the $\hat{\text{\bf{L}}}$ direction modifies the reconnection rate in a symmetric CS according to Equation \ref{eq:RecRate} \citep{cassakOtto2011}. Therefore, 
 \L-shears larger than the Alfvén speed result in a vanishing reconnection rate and the suppression magnetic reconnection in symmetric CSs with zero guide field \citet{ cassakOtto2011}. This theoretical limit to the flow shears that can support magnetic reconnection seems to be more strict in observations: we find that only 7\% of our reconnecting CSs have an \L-shear larger than 0.5, a result consistent with that of \citet{phan2020parker} and \citet{eriksson2022}.

Our observations of the normalized $\hat{\text{\bf{L}}}$-shear, show a separation between Alfvénic and reconnecting CSs at around $\Delta v_L/\Delta v_{A,L}=0.5$ (Figure \ref{fig:CSflowshear} (a)). Only 7\% of reconnecting CSs lay above this threshold, while 10\% of the Alfvénic CSs are below it. That difference in the normalized \L-shear of Alfvénic and reconnecting CSs indicates that large normalized flow-shears can suppress magnetic reconnection. 

We observe that 12\% of the Alfvénic CSs have an \L-shear larger than the Alfvénic velocity shear in the $\hat{\text{\bf{L}}}$ direction, this is in contrast with the observations of \citet{phan2020parker}, who "almost never" observed Alfvénic CSs with \L-shears that exceeded the Alfvén velocity shear. 

In panel (b) of Figure \ref{fig:CSflowshear} we show the ratio of change of the dimensional reconnection rate due to the $\hat{\text{\bf{L}}}$-shear of the CSs, $E$, with respect to the reconnection rate in the absence of a flow shear, $E_0$, according to Equation \ref{eq:RecRate}. One can observe how the reconnection rate is higher for Reconnecting CSs, than for Alfvénic CSs, as expected.

One of the mechanisms that have been proposed as drivers of magnetic reconnection is the tearing mode instability. When a CS is affected strongly enough by it, it becomes thinner and magnetic reconnection can ensue. The presence of a flow shear can affect the growth rate of the tearing mode instability, thus impacting the ability of magnetic reconnection to proceed. In panel (c) of Figure \ref{fig:CSflowshear}
we show the ratio of change of the tearing mode instability growth rate when a flow shear in the $\hat{\text{\bf{L}}}$ direction is introduced, $\gamma_{tr}/\gamma_{0,tr}$ computed according to Equation \ref{eq:TearingGrowth}, as proposed by \citet{mallet2025suppression}. For this computation we have assumed that there are only protons and no heavier ions, $Z=1$, we have assumed a temperature ratio $\tau=T_i/T_e=0.5$, which is a traditional rule of thumb approximation \citep{montgomery1972average,serbu1972explorer}. We have also assumed that the flow shear defined by \citet{mallet2025suppression} in terms of velocity and magnetic field fluctuations $\Sigma=\delta v/\delta b$ is equivalent to our definition based upon variations of the same physical properties. One can see how the growth rate of the tearing instability is consistently larger for reconnecting than for Alfvénic CSs, a result consistent with that of \citet{ervin2025impact}. In Alfvénic CSs, the growth rate of the instability is dampened, limiting the ability of the tearing instability to initiate reconnection. We can see that some Alfvénic CSs present negative values of the ratio of change of the growth rate, this can occur in the case of super-Alfvénic flow shears ($\Delta v_L>\Delta v_{A,L}$), where the tearing mode instability is quenched and reconnection is fully suppressed. 

Alfvénic CSs are shown to have consistently smaller theoretical reconnection rates \citep{cassakOtto2011} and growth rates of the tearing instability \citep{mallet2025suppression} than reconnecting CS, but can reach values of $E/E_0>0.75$ or $\gamma_{tr}/\gamma_{0,tr}>0.75$ in 10\% of the Alfvénic CS, where the reconnection rate or the tearing instability are not diminished, respectively. Our interpretation of this effect is that while large $\hat{\text\bf{L}}$-shears suppress reconnection, small $\hat{\text{\bf{L}}}$-shears do not impose it.

 Figure \ref{fig:CSflowshear} (d) shows the velocity shear in the direction of the guide field ($\hat{\text{\bf{M}}}$-shear). All CSs have $\hat{\text{\bf{M}}}$-shear less than 1, irrespective of their reconnection status, with medians 0.10 and 0.15 for Alfvénic and reconnecting CSs, respectively . \citet{liang2024impact,swisdak2026numerical} have proposed that high guide field shears can decrease the reconnection rate and increase the onset time of the linear tearing mode instability. However, higher shears than we observe are necessary for this effect to occur. Reconnection suppression via guide field directed velocity shear may be more significant in other contexts, such as the magnetopause, than in the small-scale solar wind CSs analyzed here. 

Panels (e) and (f) of Figure \ref{fig:CSflowshear} present the magnitude and orientation of the tangential flow shear $\Delta v_{LM}$. The tangential flow shear magnitude is smaller for reconnecting than for Alfvénic CSs, with median values of 16 km s$^{-1}$ and 6 km s$^{-1}$, respectively. The latter value is consistent with the median 5 km s$^{-1}$ tangential flow shear magnitude reported in \citet{eriksson2022}, for the reconnection exhausts found at 1 AU, despite the difference in the length scale of the events. The distribution of the tangential flow shear angle, $\psi$, of the reconnecting CSs  is close to uniform with a median value of 39$^{\circ}$, attributable to $L$-shear and M-shear being similarly small.  \citet{eriksson2024parker} reports a similar  distribution of the $\psi$-angle from reconnection exhaust detections at 1 AU. 
Alfvénic CSs have a considerably smaller tangential flow shear angle than reconnecting CSs, with a median of 7.1$^{\circ}$. This is consistent with the distributions of the flow shears shown in Figures \ref{fig:CSflowshear} (a) and (d), which show that the flow shear is larger in the $\hat{\textbf{L}}$ direction than in the $\hat{\textbf{M}}$ direction for Alfvénic CSs.

\begin{figure}[ht]
\epsscale{1.2}
\plotone{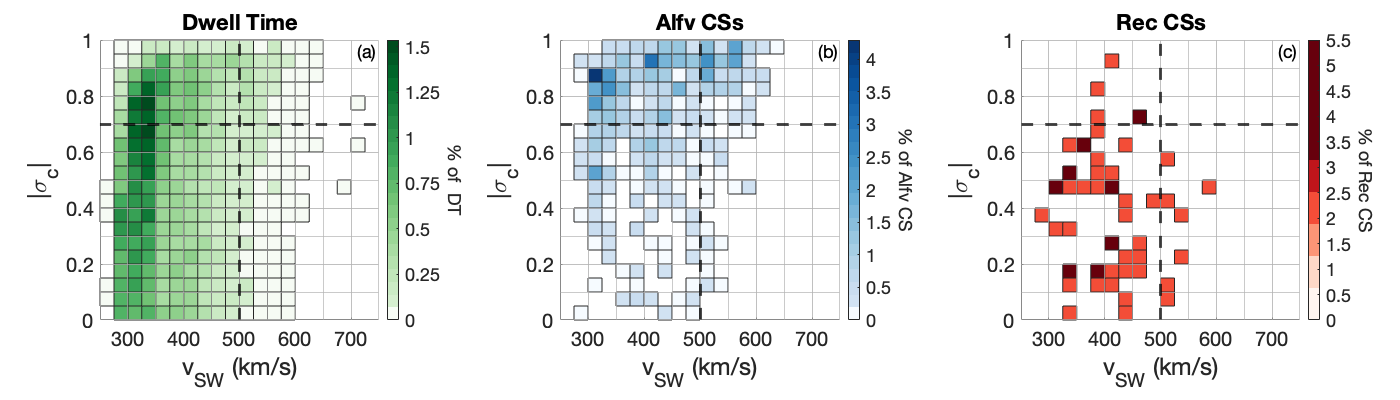}
\caption{Two dimensional distribution of solar wind speed and cross helicity in absolute value, shown for the 10 minute interval sourounding each 8 second segment of observations. Panels show: (a) the full dwell time  (green), (b) the Alvénic CSs (blue) and (c) the Reconnecting CSs (red). The vertical and horizontal dashed lines represent $v_{SW}=500$ km s$^{-1}$ and $|\sigma_c|=0.7$ for visual aid.  
\label{fig:CSCrossHel}}
\end{figure}

 Figure \ref{fig:CSCrossHel} shows the average solar wind speed and the absolute value of the cross helicity computed in the 10 minutes that surround each segment of 8 seconds of the Dwell Time. The cross-helicity is computed as  
$$\sigma_c=\frac{2\langle \delta\bf{v}\cdot\delta\bf{v_A}\rangle}{\langle\delta\bf{v}^2\rangle+\langle\delta\bf{v_A}^2\rangle}$$

Each data segment of the Dwell Time that is longer than 8 s is divided in 8 s sub-segments. In this way, all the Dwell Time is composed of 8s data pieces. The cross-helicity is computed for each of these data pieces by using the magnetic field, ion density, and ion velocity data collected in the 10 minutes lapse surrounding the central time of each of the 8s data pieces. These data are resampled to the 4s cadence at which the normal mode velocity data are collected.

Panels (b) and (c) of Figure \ref{fig:CSCrossHel} show that 70\% of the Alfvénic CSs reported in this study are detected at time intervals of Alfvénic solar wind, i.e., intervals where the cross helicity is $|\sigma_c|>0.7$ while only 9\% of the Reconnecting CSs are detected under the same circumstances. This is consistent with the findings of \citet{eriksson2024parker,ervin2025impact}. Reconnection events are found more frequently in non-Alfvénic solar wind, supporting the idea that Alfvénic flows can suppress magnetic reconnection \citep{mallet2025suppression}.

\subsection{Current sheet planarity}\label{sec:flatness}

We now turn our attention to the magnetic structure of these CSs. In the interest of clarity, let us remind the reader that $\hat{\textbf{l}}$, $\hat{\textbf{m}}$, $\hat{\textbf{n}}$ refer to the reconnection, guide field, and normal directions of a CS, respectively, while $\hat{\text{\bf{l}}}_{MVA}$, $\hat{\text{\bf{m}}}_{MVA}$, $\hat{\text{\bf{n}}}_{MVA}$ refer to those same directions computed using the classic MVA method, and $\hat{\textbf{L}}$,  $\hat{\textbf{M}}$, $\hat{\textbf{N}}$ are again the same directions, computed using the hybrid-MVA method. 

 A CS that is planar, i. e., that has a 1D geometry is defined by a stable, non-fluctuating $\hat{\textbf{n}}$ direction, which means that $\partial B_n/\partial n = 0$. Overall, $\nabla\cdot \bf{B}$$=\frac{\partial B_l}{\partial l}+\frac{\partial B_m}{\partial m}+\frac{\partial B_n}{\partial n}=0$ must be satisfied for any magnetic field configuration. Therefore, for a planar CS, $\partial B_l/\partial l+ \partial B_m/\partial m \simeq 0$. Classic MVA finds the $\hat{\textbf{n}}$ direction that minimizes the derivative of $B_n$ in the $\hat{\text{\bf{n}}}_{MVA}$ direction, while Hybrid MVA forces $B_N = 0$ at the two edges of the CS. $B_N$ may indeed be non-zero inside the CS, but we observe that it is generally small: in 91\% of them $\langle B_N\rangle $ accounts for less than 5\% of the full $|\mathbf{B}|$. ºIn order to assess the planarity of the CSs, we compare the normal directions obtained from classic and hybrid-MVA: $\hat{\text{\bf{n}}}_{MVA}$ and $\hat{\text{\bf{N}}}$, respectively. In non-ideal situations, the magnetic field rotation is not contained within a plane, $\frac{\partial B_n}{\partial n}\neq 0$, and there is no unique normal direction across the full CS, meaning that the CS is not planar. 

\begin{figure*}[b]
\includegraphics[width=\textwidth]{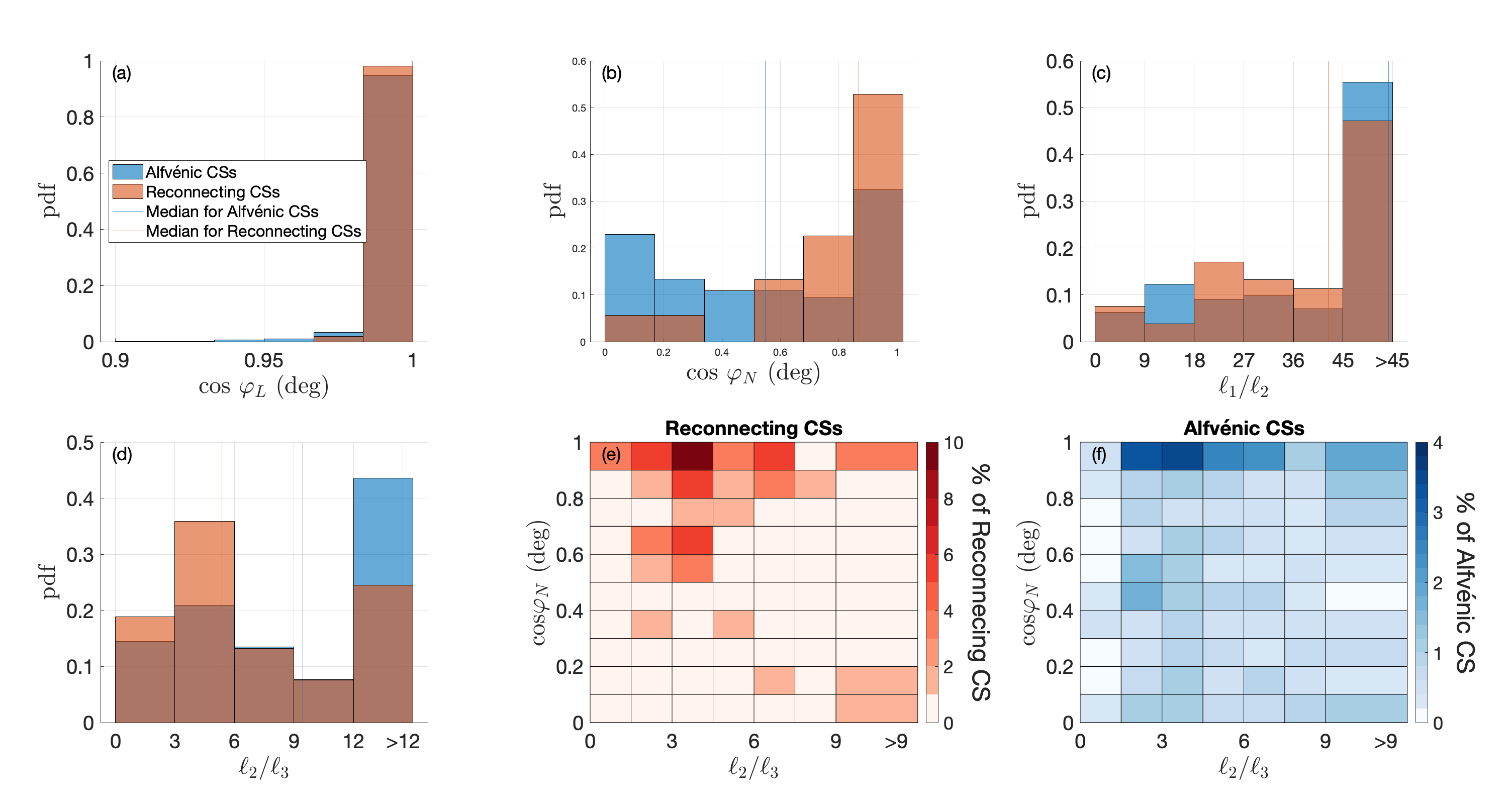}
\caption{CS planarity. Panel (a) shows the distribution of the cosine of the angles formed by the $\hat{\text{\bf{L}}}$ and $\hat{\text{\bf{l}}}_{MVA}$ vectors for Alfvénic (blue) and Reconnecting (red) CSs. Panel (b) shows the distribution formed by the $\hat{\text{\bf{N}}}$ and $\hat{\text{\bf{n}}}_{MVA}$ vectors, . Panel (c) shows the ratio between the $\ell_1$ and $\ell_2$ eigenvalues yielded by classic MVA for Alfvénic (blue) and Reconnecting (red) CSs. Panel (d) shows the ratio between the $\ell_2$ and $\ell_3$ eigenvectors obtained in the same manner. Panels (e) and (f) show the bivariate distribution of the variables shown in panels (b) and (d) for Reconnecting and Alfvénic CSs respectively.
\label{fig:CSFlat}}

\end{figure*}

The effect of the variability of $B_n$ is taken into account by classic-MVA, since this method finds the $\hat{\text{\bf{n}}}_{MVA}$ direction as an average direction that minimizes $\frac{\partial B_n}{\partial n}$. In contrast, hybrid-MVA disregards the variability of $B_n$ and produces the normal vector that satisfies $B_N=0$ at the CS borders, assuming the CS is planar and that $B_N$ does not vary across the CS. Thus, in order to further investigate the effect of the variation of $B_N$ on the planarity of CSs, we compare the vectors $\hat{\text{\bf{l}}}$ and $\hat{\text{\bf{n}}}$ obtained by both methods. The comparison is performed by computing the angle the two vectors form, $\cos\varphi_L = \hat{\text{\bf{l}}}_{MVA}\cdot\hat{\text{\bf{L}}}$ and $cos\varphi_N = \hat{\text{\bf{n}}}_{MVA}\cdot\hat{\text{\bf{N}}}$ (shown in panels (a) and (b) of Figure \ref{fig:CSFlat}, respectively).

Figure \ref{fig:CSFlat}(a) shows that the cosine of $\varphi_L$ is close to 1 for all the CSs, which means that the $\hat{\textbf{l}}$ directions in both coordinate systems agree well. The cosine of $\varphi_N$ is shown in panel (b). The amount of reconnecting CSs increases as $\cos\varphi_N$ increases, while the  distribution of the Alfvénic CSs is "U"-shaped, with maxima at 0 and 1. This means that  the $\hat{\text{\bf{N}}}$ directions provided by the two methods are in agreement in most reconnecting CSs. Thus, the impact of $\frac{\partial B_N}{\partial N}$ being non-zero, i.e., the impact the magnetic field rotation wandering out of the plane defined by $\hat{\text{\bf{N}}}$, is small, so we can consider $\frac{\partial B_N}{\partial N}$ itself to be small. That can be interpreted as CSs with larger  $\cos\varphi_N$ being more planar, in the sense of the current layer being contained in a flat plane that is reasonably well defined, and  the assumption of planarity being meaningful.

To assess planarity, we also look at the eigenvalue ratios $\ell_1/\ell_2$ and $\ell_2/\ell_3$ provided by classic MVA (panels (c) and (d) in Figure \ref{fig:CSFlat}, respectively), in which the last quantity has usually been interpreted as a proxy for the validity of the planarity assumption. We do not find that $\ell_1/\ell_2$ is significantly larger for reconnecting versus Alfvénic CSs; in contrast, $\ell_2/\ell_3$ reaches higher values for Alfvénic CSs, which would at first sight imply that Alfvénic CSs are somewhat more planar than Reconnecting ones. However, neither of these parameters is a straightforward measure of the planarity of the structure, but instead a comparison between the variations of the magnetic field in two of the directions. In particular, $\ell_1/\ell_2$ measures how much more variable the magnetic field is in the $\hat{\text{\bf{l}}}_{MVA}$ than in the $\hat{\text{\bf{m}}}_{MVA}$ direction, while $\ell_2/\ell_3$ gives an analogous comparison between $\hat{\text{\bf{m}}}_{MVA}$ and $\hat{\text{\bf{n}}}_{MVA}$, i.e., $\ell_1/\ell_2\propto\frac{\partial B_{l_{MVA}}}{\partial n_{MVA}}/\frac{\partial B_{m_{MVA}}}{\partial n_{MVA}}$ and $\ell_2/\ell_3\propto\frac{\partial B_{l_{MVA}}}{\partial n_{MVA}}/\frac{\partial B_{m_{MVA}}}{\partial n_{MVA}}$. The high $\ell_1/\ell_2$ values observed therefore mean that the $\hat{\text{\bf{l}}}_{MVA}$ direction, which is very similar to the $\hat{\text{\bf{L}}}$ direction by virtue of Figure \ref{fig:CSFlat}(a), corresponds to a magnetic field variation that is significantly larger than the other directions, for both Reconnecting and Alfvénic CSs, as expected. On the other hand, 55\% of Reconnecting CSs and 35\% of Alfvénic CSs have $\ell_2/\ell_3<6$. For these CSs, the variations of the magnetic field in the $\hat{\text{\bf{m}}}_{MVA}$ and $\hat{\text{\bf{n}}}_{MVA}$ are of the same order of magnitude, and thus MVA is less accurate in assigning one direction or the other as the direction of minimum variation. High values of $\ell_2/\ell_3$ would imply that the direction of minimum variation ($\hat{\text{\bf{n}}}_{MVA}$) can be reliably chosen, thus neatly defining the plane of the CS discontinuity. However, low values of the same parameter $\ell_2/\ell_3$ do not necessarily mean the opposite. The reason for this is twofold: firstly, $\ell_2/\ell_3 \sim 1$ does not preclude the magnetic field variations in the ($\hat{\text{\bf{n}}}_{MVA}$) direction from being quite small and thus defining a constant $B_{n_{MVA}}$ along the CS; secondly, $\ell_2/\ell_3$ may be close to one due to the encroaching of magnetic field fluctuations related to other physical processes, producing corrugations of the magnetic field in an otherwise well-defined overall normal direction.

Then, CSs that present high $\cos{\varphi_N}$ and high $\ell_2/\ell_3$, located at the top right of panels (e) and (f) of Figure \ref{fig:CSFlat}, can be interpreted as planar. For these CSs, the $\hat{\text{\bf{n}}}_{MVA}$ and $\hat{\text{\bf{N}}}$ directions defined by classic and hybrid-MVA coincide, which we see as a first sign of the CS being planar. The large $\ell_2/\ell_3$ value signals that $\frac{\partial B_{n_{MVA}}}{\partial n_{MVA}}$ is small, which also supports the planar interpretation.

When either $\cos{\varphi_N}$ or $\ell_2/\ell_3$ become small, the interpretation becomes less clear. Even for CSs with $\cos\varphi_N$ close to 0, $\cos\varphi_L$ is close to 1, which means that classic MVA and hybrid-MVA interchange the normal, $\hat{\text{\bf{n}}}$, and guide field, $\hat{\text{\bf{m}}}$, directions. The discrepancy may be due to the CS being extremely planar, with small fluctuations in both $\hat{\text{\bf{m}}}$ and $\hat{\text{\bf{n}}}$ directions, or to the structure being completely non planar, with large magnetic field fluctuations compromising the performance of the classic and MVA methods. A low $\ell_2/\ell_3$ value indicates fluctuations that are similar in order of magnitude in the two directions $\hat{\text{\bf{m}}}_{MVA}$ and $\hat{\text{\bf{n}}}_{MVA}$

Recent studies \citep{wang2024} have shown that classic MVA is particularly inaccurate in the solar wind due to its anisotropy, which makes fluctuations of the magnetic field in the direction parallel to the background magnetic field smaller than fluctuations in the perpendicular directions. In the case of the CSs studied here, the guide field direction is mostly aligned with the background magnetic field, and therefore fluctuations in the guide field direction tend to be smaller, due to the anisotropy of the solar wind fluctuations. The smaller amplitude of these fluctuations makes MVA more likely to identify the direction of the guide field as the direction of minimum variation, $\hat{\text{\bf{n}}}_{MVA}$. 

We observe that 21\% of Alfvénic and 26\% of Reconnecting CSs exhibit $\cos\varphi_N>2/3$ and $\ell_2/\ell_3>9$. We believe these CSs are likely planar, while the planarity of the rest of the CSs in the catalog cannot be assessed with a single-spacecraft observation. 




%


\section{Conclusions}\label{conclusion}
In this article we present a statistical study of the properties of small-scale solar wind magnetic reconnection and CSs. These structures were observed by Solar Orbiter and sampled at some of the smallest resolutions yet reached in the solar wind. We run the algorithm described in Section 3 of the companion paper \citep{Albert2026P1} on 103 h of high resolution ion and magnetic field data and automatically find 974 CSs of thicknesses in the range 0.06 $\delta_i \le \lambda \le$ 86.4 $\delta_i$, classified as Reconnecting (53), Alfvénic (838) and Undetermined (83). On this range of CS thickness we find 0.51 reconnection exhausts per hour of observation, and 9.4 CS h$^{-1}$. 
We studied the properties of current sheets near ion length-scales in the solar wind (0.3 - 0.9 AU) using in-situ, high-resolution Solar Orbiter measurements. We assessed the occurrence of magnetic reconnection and the mechanisms that may influence this occurrence, to gain insight on the role of magnetic reconnection in dissipating magnetic energy near ion length-scales. This analysis yielded several conclusions:

\begin{itemize}

    \item The orientation of the solar wind small-scale CSs studied in this catalog appears to be isotropic, with no visible signatures to the contrary. This trend is followed by CSs of all thicknesses in the studied range and by both reconnecting and Alfvénic CSs.
    
    \item The classification of the CSs in our catalog by the type of discontinuity that describes them is highly inconclusive: only 7.3\% of CSs are classified as TDs, while the remaining CSs are not classified as RDs and remain unclassified. The criteria used are insufficient for robust classification, partly because the use of hybrid MVA makes the requirement of constant and non-zero $B_N$ for RD classification mostly unattainable.

    \item The median normalized jet velocity in reconnecting CSs with respect to the theoretical outflow velocity is 0.71.

    \item The suppression of magnetic reconnection via diamagnetic effect \citep{swisdak2003diamagnetic,swisdak2010vector} is not important for ion-scale CSs in the slow solar wind.

    \item The vast majority of the CSs found in this catalog are Alfvénic (86\%), which is consistent with the  large cross-helicity of the sampled solar wind. These Alfvénic CSs generally present large \L-
    shears, which prevent them from reconnecting. Only the small fraction of non-Alfvénic CSs present small \L-shears and can reconnect. Flow shear in the \L direction appears to be a major variable influencing the distinction between reconnecting and non-reconnecting CSs, with large \L-shears suppressing magnetic reconnection at ion scales in the Alfvénic solar wind.
    
    \item 
    Alfvénic CSs show consistently larger normalized \L-shears than reconnecting CSs, with an apparent separation between the two populations at about $\Delta v_L/\Delta v_{A,L} = 0.5$. 93\% of Reconnecting CSs have \L-shears below that threshold, while 90\% of Alfvénic CSs are above it. This observation threshold is smaller than predicted by theory \citep{cassakOtto2011}.

    \item The velocity shear in the $\hat{\text{\bf{M}}}$ direction of the detected ion-scale solar wind CSs is too small to suppress magnetic reconnection.

    \item We study the planarity of the CS by looking at the ratio between  the second and third eigenvalues provided by classic MVA, $\ell_2/\ell_3$ and at the angle formed by the CS normal direction as computed by hybrid and classic MVA, $\varphi_N$. We conclude that for 26\% of Alfvénic and 21\% of Reconnecting CSs the combination of high $\ell_2/\ell_3$ and high $\cos\varphi_N$ means that these events are planar. Planarity of the structure cannot be guaranteed for the rest of the CSs.

\end{itemize}

In conclusion, Solar Orbiter has enabled us to study the properties of CSs and discontinuities in the pristine solar wind at small-scales. We have been able to show that the properties of the ambient solar wind can control the ability of CSs to undergo magnetic reconnection. Large flow shears in the \L direction, linked to a high cross helicity of the ambient solar, suppress magnetic reconnection. . We have assessed the planarity of small-scale CSs based on the available single spacecraft data, and concluded that most CSs are likely non planar, with various levels of confidence.  Therefore, future multi-spacecraft
measurements in the solar wind will be invaluable in better understanding these structures.

\section{Acknowledgements}

We gratefully acknowledge the work done by all the members of the international team that made the Solar Orbiter mission possible and successful. In particular, we would like to thank the teams in charge of the Magnetometer and Solar Wind Analyser instruments on board Solar Orbiter.
Solar Orbiter is an international cooperation mission between ESA and NASA, operated by ESA. 
We acknowledge support from Grant PID2023-147331OB-I00 funded by MICIU/AEI/ 10.13039/501100011033 and FEDER, UE, and Seneca Agency from
Region of Murcia (Grant 21910/PI/22). 
Victor Montagud-Camps is supported by the Vicenç Mut grant POSTDOC2024\_58 funded by Conselleria d'Educació i Universitats del Govern de les Illes Balears i pel Fons Social Europeu +".

\bibliography{Biblio}{}
\bibliographystyle{aasjournal}

\end{document}